\documentclass[a4paper, 10pt]{article}

\usepackage{amsmath}
\numberwithin{equation}{section} % Numbers equation as 1.1 instead of just 1,2,3,... as it is standard in the article class. The command is provided by the amsmath package

\usepackage{amsfonts}
\usepackage{amssymb}
\usepackage{slashed, cancel}
\usepackage{geometry}
\usepackage[english]{babel}
\usepackage[utf8]{inputenc}
\usepackage{graphicx,url}
\usepackage{color}
\usepackage[dvipsnames]{xcolor}
\usepackage[colorlinks,
linkcolor=NavyBlue, urlcolor=DarkOrchid, citecolor=ForestGreen,
pagebackref=true,linktocpage=true,
hyperfootnotes=true,
pdfstartview=FitH,bookmarksopen,bookmarksnumbered]{hyperref}
\usepackage{cite}
\usepackage[font=footnotesize,labelfont=bf]{caption}

\usepackage{tikz,pgfplots}
\usetikzlibrary{arrows,shapes,trees,calc,positioning,patterns}

\usepackage{array} % To modify spacing inside tables

\usepackage{mathrsfs} % Alternative calligraphic font in math mode. Used for the Lagrangian symbol.

\usepackage{feynmp}
\newcommand{\Lagr}{\mathscr{L}}
\newcommand{\GeV}{{\rm \,GeV}}
\newcommand{\TeV}{{\rm \,TeV}}
\newcommand{\ifb}{\,{\rm fb}^{-1}}%  Inverse femtobarns.
\def\ie{\textit{i.e. }}
\def\eg{\textit{e.g. }}
\def\etc{\textit{etc. }}
\def\pT{p_{\rm T}}
\def\mX{m_{\chi}}
\def\gX{g_{\chi}}
\newcommand{\met}{\slashed{E}_T}

\newcommand{\mZp}{m_{Z'}}
\newcommand{\Mmed}{M_{\rm med}}
\newcommand{\mmed}{m_\textrm{med}}

\newcommand\blfootnote[1]{%
  \begingroup
  \renewcommand\thefootnote{}\footnote{#1}%
  \addtocounter{footnote}{-1}%
  \endgroup
}

\begin{document}
%\vspace*{4cm}
\title{Simplified Dark Matter Models}

\author{Enrico Morgante}

%\address{DESY, Notkestra{\ss}e 85, D-22607 Hamburg, Germany}

\begin{flushright} 
DESY 18-047
\end{flushright}

\vspace{2cm}

\begin{center}
\textbf{\Large Simplified Dark Matter Models} \\[.5cm]
\large Enrico Morgante \\[.3cm]
\textit{\small Deutsches Elektronen-Synchrotron DESY, Notkestra{\ss}e 85, D-22607 Hamburg, Germany} \\
\texttt{\small enrico.morgante@desy.de}
\end{center}

\vspace{1cm}

\textbf{Abstract:} I review the construction of Simplified Models for Dark Matter searches. After discussing the philosophy and some simple examples, I turn the attention to the aspect of the theoretical consistency and to the implications of the necessary extensions of these models.%
\blfootnote{The author declares that there is no conflict of interest regarding the publication of this paper.}

\vspace{.5cm}

\tableofcontents

\newpage

\section{Introduction}

Producing and studying the properties of the Dark Matter particles (DM) at the LHC is an extremely exciting possibility, that would open the door to a new understanding of the interplay between astrophysics, cosmology and particle physics.
Essentially all the naturalness-inspired scenarios can accommodate the presence of a good Dark Matter candidate: a neutral and very long-lived particle that was copiously produced in the early universe and then lost thermal contact with the SM (if it ever occurred) leaving a relic density $\Omega_{\rm DM}\sim 0.26$ of cold particles. The fact that such a stable weakly interacting massive particle with a mass around the weak scale has automatically a relic abundance close to the measured one is a remarkable property which is often dubbed \emph{WIMP miracle}.
The LHC is a perfectly suited machine to look for this kind of particles, and current bounds from ATLAS and CMS complement those from direct and indirect searches.

A key task in these studies is that of choosing a theoretical framework to compare with data and compare the results of different experiments.
Given the plethora of particle physics models beyond the SM providing a WIMP candidate, it is highly desirable to study the signatures of this DM candidate in a model-independent way.
In the early stages of the LHC, this was achieved by means of the effective field theory approach (EFT). In this framework, the Standard Model (SM) is complemented by a set of non renormalizable operators, that parametrize the interaction of the DM particle with SM fields in terms of one effective scale $\Lambda$, and of the DM mass $\mX$~\cite{Goodman:2010ku}.
The EFT approach has proven to be very useful in the analysis of LHC Run I data \cite{Beltran:2010ww, Goodman:2010yf, Bai:2010hh, Goodman:2010ku, Fox:2011fx, Rajaraman:2011wf, Fox:2011pm, Shoemaker:2011vi, Cotta:2012nj, Dreiner:2012xm, Chae:2012bq, Fox:2012ru, DeSimone:2013gj, Dreiner:2013vla, Chen:2013gya},
because of the great advantage of giving bounds that are as model-independent as possible: for a given choice of the spin of the DM particle, the number of operators that can couple it to the SM and may give interesting signatures at the LHC is limited, for a fixed mass dimension.
Since direct and indirect detection of WIMPs, as well as WIMP production at the LHC, all require an interaction of the WIMPs with the SM particles, and such an interaction may be generated by the same operator, the EFT approach has the additional advantage of facilitating the analysis of the correlations between the various kinds  of experiments.

The important drawback of the EFT description is its intrinsic energy limitation. At energies larger than some cutoff $\Lambda$, the contribution of higher dimension operators to the computation of scattering amplitudes becomes comparable to the lower ones, signalling the breakdown of perturbativity.
More in particular, if the EFT is seen as the low energy limit of a theory with a mediator of mass $M$ which is above the energy scale probed by the experiment, the cutoff is obtained as $\Lambda^2\sim M^2/g^2$, where $g$ is some combination of the coupling constants, and the theory is valid up to a momentum exchange $p^2\lesssim M^2 \sim \Lambda^2$, where we have assumed $g\sim 1$.
In Direct and Indirect detection this constraint is typically satisfied thanks to the low velocity of the incoming particle. On the other hand, the momentum exchanged in the partonic interactions at the LHC is of order few TeV, larger than the values of $\Lambda$ that can be excluded within the EFT framework, making the na\"ive EFT bounds unreliable except for values of the couplings close to the perturbative bound $g\lesssim 4\pi$~\cite{Busoni:2013lha,Busoni:2014sya,Busoni:2014haa}.
In principle this does not mean that the EFT approach is not useful.
Recasting procedures can be adopted to re-derive bounds considering only a fraction of the events in the simulation that correspond to those which fulfil the requirement on the momentum~\cite{Busoni:2014sya,Racco:2015dxa}.
Clearly, the new bounds would be much weaker, but their simplicity still suggests that they should not be disregarded.

Partly in response to the problems of EFTs, and partly inspired by their rich phenomenological implications, in more recent years the LHC community has turned its attention to the tool-kit of simplified models.
Such models are characterized by the most important state mediating the interaction of the DM particle with the SM, as well as the DM particle itself
(see for example~
\cite{Dudas:2009uq, Goodman:2011jq, An:2012va, Frandsen:2012rk, Dreiner:2013vla, Cotta:2013jna}
for early proposals).
Including the effect of the mediator's propagator allows to avoid the energy limitation of the EFT, and simplified models are able to describe correctly the full kinematics of DM production at the LHC, at the price of a moderately increased number of parameters.
As we are going to discuss below, the introduction of simplified models opens a new set of possibilities compared to the simpler EFT approach, while opening at the same time a set of new questions.

This paper is structured as follows. In Section~\ref{sec:philosophy} we are first going to describe the construction of DM simplified models from a bottom-up approach, providing some examples in Section~\ref{sec: simp mods}.
Then, in Section~\ref{sec:simpmodsdiscussion}, we will point out the theoretical issues of such a construction, introducing a second generation of simplified models that have gained a lot of attention in recent times.
Section~\ref{sec:conclusions} will contain our conclusions.

Thorough discussions about simplified DM models may be found in~\cite{Abdallah:2014hon, Malik:2014ggr, Abdallah:2015ter, Abercrombie:2015wmb, DeSimone:2016fbz, Arcadi:2017kky}. The second generation models of Section~\ref{sec:simpmodsdiscussion} are discussed in~\cite{Bauer:2016gys, Kahlhoefer:2017dnp}.

\section{Philosophy of simplified DM models}
\label{sec:philosophy}

As in the case of the EFT, the idea behind simplified models is to provide a good representation of possibly all realistic WIMP scenarios within the energy reach of the LHC, restricting to the smallest possible set of benchmark models, each with the minimal number of free parameters.
Simplified models should be complete enough to give an accurate description of the physics at the scale probed by colliders, but at the same time they must have a limited number of new states and parameters.

The starting point is always the SM Lagrangian, complemented with a DM particle and a mediator that couples to it, through renormalizable operators, to quarks and gluons, which is necessary for the production of these states at a hadron collider. A coupling to other SM particles can be included as well and will add interesting experimental signatures to the model.
In general, some simplifying assumptions can be made: for example, one can take all couplings to be equal, or the couplings to third generation's quarks to be dominant.
Interactions that violate the accidental global symmetries of the SM must be handled with great care.
Indeed, constraints on processes that violate these symmetries are typically very strong, and may overcome those coming from DM searches or even rule out all of the interesting parameter space of the simplified model.
For this reason, CP, lepton number and baryon number conservation is typically assumed, together with Minimal Flavour Violation (MFV).\footnote{
Constraints on BSM models from CP and flavour violating observables are very strong, and the energy scale at which new physics may show up must be larger than tens of TeV in the best case, if the flavour structure of the model is generic.
Minimal Flavour Violation is a way to reconcile these constraints with possible new physics at the TeV scale \cite{DAmbrosio:2002vsn}.
The basic idea is that the structure of flavour changing interactions must reproduce that of the SM.
The SM is invariant under the flavour group $\mathcal{G}_F = {\rm SU}(3)_q\times {\rm SU}(3)_u \times {\rm SU}(3)_d$, except from a small breaking associated to the Yukawa matrices $Y_u$ and $Y_d$.
The invariance is restored if these matrices are regarded as ``spurions'' with transformation law $Y_u\sim(3,\bar 3,1)$ and $Y_d\sim(3,1,\bar 3)$.
Imposing MFV amounts to requiring that new physics is invariant under $\mathcal{G}_F$.}
Even with this assumption, there are cases in which constraints from flavour physics may be stronger than those coming from mono-X searches \cite{Dolan:2014ska} (see also \cite{Agrawal:2014aoa} for a discussion of a non-minimally flavour violating dark sector).

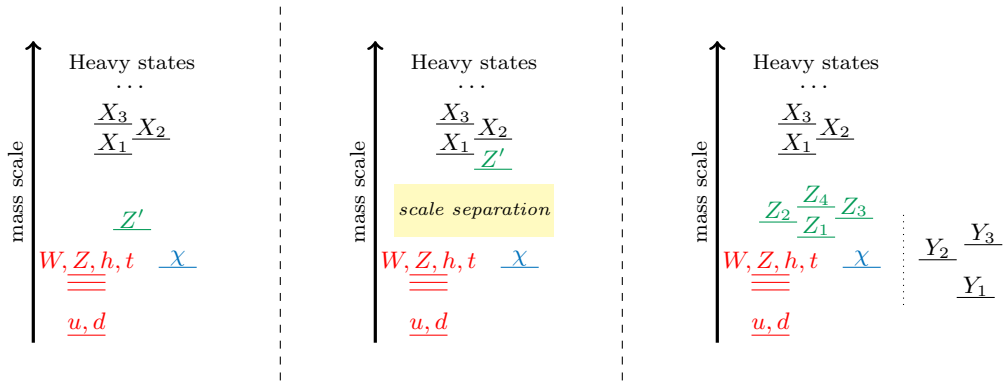
\begin{figure}[t!]
\centering
\begin{tikzpicture}
\tikzset{
   spectrum/.pic={
   \draw[very thick, ->] (-.25,0) -- ++(0,4)
	node[pos=.5,sloped,above,align=center] {\footnotesize mass scale};
\draw[red] (.2,.1) -- ++(.5,0)
	node[above=-1mm, red, pos=.5, align=center] {\small $u,d$};
\draw[red] (.2,.7) -- ++(.5,0);
\draw[red] (.2,.8) -- ++(.5,0);
\draw[red] (.2,.9) -- ++(.5,0)
	node[above=-1mm, red, pos=.5, align=center] {\small $W,Z,h,t$};
%\draw[ForestGreen] (.8,1.5) -- ++(.5,0)
%	node[above=-1mm, ForestGreen, pos=.5, align=center] {\small $Z'$};
\draw[RoyalBlue] (1.4,1) -- ++(.5,0)
	node[above=-1mm, RoyalBlue, pos=.5, align=center] {\small $\chi$};
\draw[black] (.55,2.5) -- ++(.5,0)
	node[above=-1mm, black, pos=.5, align=center] {\small $X_1$};
\draw[black] (1.05,2.7) -- ++(.5,0)
	node[above=-1mm, black, pos=.5, align=center] {\small $X_2$};
\draw[black] (.55,2.9) -- ++(.5,0)
	node[above=-1mm, black, pos=.5, align=center] {\small $X_3$};
\node[black, align=center] at (1.05,3.4) {\small \ldots};
\node[black, align=center] at (1.05,3.7) {\footnotesize Heavy states};
	}
}

\pic at (0,0) {spectrum};
\draw[ForestGreen] (.8,1.5) -- ++(.5,0)%
	node[above=-1mm, ForestGreen, pos=.5, align=center] {\small $Z'$};
\draw[black, dashed] (3,-.5) -- ++(0,5);

\begin{scope}[shift={(4.5,0)}]
\pic at (0,0) {spectrum};
\draw[ForestGreen] (1.05,2.3) -- ++(.5,0)%
	node[above=-1mm, ForestGreen, pos=.5, align=center] {\small $Z'$};
\fill [yellow!30] (0,1.4) rectangle ++(2.1,.7)
	node[pos=.5, black] {\scriptsize \it scale separation};
\draw[black, dashed] (3,-.5) -- ++(0,5);
\end{scope}

\begin{scope}[shift={(9,0)}]
\pic at (0,0) {spectrum};

\draw[ForestGreen] (.8,1.4) -- ++(.5,0)%
	node[above=-1mm, ForestGreen, pos=.5, align=center] {\small $Z_1$};
\draw[ForestGreen] (.3,1.6) -- ++(.5,0)%
	node[above=-1mm, ForestGreen, pos=.5, align=center] {\small $Z_2$};
\draw[ForestGreen] (1.3,1.65) -- ++(.5,0)%
	node[above=-1mm, ForestGreen, pos=.5, align=center] {\small $Z_3$};
\draw[ForestGreen] (.8,1.8) -- ++(.5,0)%
	node[above=-1mm, ForestGreen, pos=.5, align=center] {\small $Z_4$};

\draw[dotted] (2.2,.5) -- ++(0,1.2);
\draw[black] (2.9,.6) -- ++(.5,0)%
	node[above=-1mm, black, pos=.5, align=center] {\small $Y_1$};
\draw[black] (2.4,1.1) -- ++(.5,0)%
	node[above=-1mm, black, pos=.5, align=center] {\small $Y_2$};
\draw[black] (3,1.3) -- ++(.5,0)%
	node[above=-1mm, black, pos=.5, align=center] {\small $Y_3$};
\end{scope}

\end{tikzpicture}

\caption{\textit{Left:} a simplified model viewed as a sector of a more general new physics scenario. The SM is complemented by the DM particle $\chi$ and a mediator $Z'$.
Other heavy states $X_1, X_2, X_3,\ldots$ may be integrated out because they are very heavy.
\textit{Centre:} in the case in which the mediator $Z'$ itself has a very large mass, it may be integrated out as well and the interaction is mediated by effective operators.
\textit{Right:} In a generic setup the DM-SM interaction is mediated by a number of operators, possibly interfering with each other. Moreover, additional states $Y_1,Y_2,Y_3,\ldots$ that do not couple directly to DM may be present. Those states can constitute a new handle on the dark sector, other than DM. The `less simplified' models described in section~\ref{sec:simpmodsdiscussion} fall in this category.}
\label{fig:simpmodspectrum}

\end{figure}

Most simplified models of interest may be understood as the limit of a more general new-physics scenario, where all new states but a few are integrated out because they have a mass larger than the energy scale reachable at the LHC or because they have no role in DM interactions with the SM.
Similarly, in the limit where the mass of the mediator is very large, the EFT framework may be recovered by integrating out the mediator.
On the contrary, there are new physics models which cannot be recast in terms of vanilla simplified models, typically because more than just one operator is active at the same time, possibly interfering with each other.
The situation is summarized in Fig.~\ref{fig:simpmodspectrum}.

Even if this may sound obvious, we should stress that the correspondence between simplified models and EFT is not one to one.
Simplified models that involve mediators of different spin nature may give rise to the same effective operator after a Fierz rotation, as pointed out in~\cite{Racco:2015dxa} with the example of a Majorana DM particle embedded in a $Z'$ model or a SUSY-inspired model with coloured scalar mediators in the t-channel.

Even when a simple correspondence between the EFT and the simplified model is assumed, limits on the EFT can not be readily translated onto the simplified model because of the possible resonant enhancement (that would make the limit stronger) or the typically softer missing energy spectrum (that would weaken the limit)~\cite{Buchmueller:2013dya}. Moreover, the different missing energy spectrum may require \textit{ad hoc} optimization strategy by the experimental searches, and considering mediators of different mass requires different optimization for each case.
Finally, models with a heavy mediator that would correspond to the EFT limit tend to predict a too large relic DM density (assuming no deviation from the standard cosmological history and no states other than the SM ones to annihilate into) and are therefore less appealing as a model of DM~\cite{Busoni:2014gta}.

From the point of view of LHC searches, the enlarged physical spectrum and parameter space of simplified models with respect to the EFT represents a challenge, and a greater variety of search channels is involved.
While within the EFT approach the mono-X searches hold the stage, simplified models of DM can be constrained also with multi-jet + MET searches, with di-jet and di-leptons resonance searches and many others, depending on the degree of sophistication and on the ingredients of the model. Interestingly, many of these searches do not involve the DM particle, but only the mediator, and constraints are often stronger than the mono-X ones.
On the other hand, the EFT is still a useful tool when dealing with strongly coupled theories, where a description in terms of a perturbative simplified model is not viable~\cite{Bruggisser:2016ixa,Bruggisser:2016nzw}.

\section{First generation of simplified models}
\label{sec: simp mods}

\subsection{$s$-channel mediators}

We are now going to list a few examples of the simplified models of relevance for LHC searches, starting with those that include a fermionic DM $\chi$ (which for now we assume to be a Dirac spinor, but this is not necessary) and a mediator exchanged in the $s$-channel.
The models under consideration are the following:
\begin{align}
\Lagr_V		& \supset 	\frac{1}{2}m_V^2 V_\mu V^\mu -m_\chi\bar\chi\chi -g_\chi V_\mu\bar\chi\gamma^\mu\chi -g_q^{ij}V_\mu\bar q_i\gamma^\mu q_j\,, \label{eq:V} \\
\Lagr_A	& \supset 	\frac{1}{2}m_A^2 A_\mu A^\mu -m_\chi\bar\chi\chi -g_\chi A_\mu\bar\chi\gamma^\mu\gamma_5\chi -g_q^{ij} A_\mu\bar q_i\gamma^\mu\gamma_5q_j \label{eq:A}\,,
\end{align}
for a spin-1 mediator and
\begin{align}
\Lagr_S	& \supset	-\frac{1}{2}m_S^2S^2 -m_\chi\bar\chi\chi -y_{\chi}S\bar{\chi}\chi -y_q^{ij}S\bar{q}_iq_j+{\rm h.c.} \label{eq:S} \,, \\
\Lagr_P	& \supset	-\frac{1}{2}m_P^2P^2 -m_\chi\bar\chi\chi -i y_{\chi}P\bar{\chi}\gamma_5\chi -i y_q^{ij}P\bar{q}_i\gamma_5 q_j+{\rm h.c.} \label{eq:P} \,.
\end{align}
for a spin-0 mediator, where $V,A,S,P$ stand for a vector, axial-vector, scalar, or a pseudo-scalar mediator respectively, $q=u,d$ and $i,j=1,2,3$ are flavour indices.
In the heavy $\mmed$ limit, the mediators can be integrated out, recovering the effective operators
\begin{equation}
\begin{aligned}
\mathrm{D1:}\,\, & \frac{m_q}{\Lambda^3} \bar\chi  \chi\;\bar q  q \qquad &
\mathrm{D4:}\,\, & \frac{m_q}{\Lambda^3} \bar\chi \gamma^5 \chi\;\bar q \gamma^5 q \\
\mathrm{D5:}\,\, & \frac{1}{\Lambda^2} \bar\chi \gamma_\mu \chi\;\bar q \gamma_\mu q \qquad &
\mathrm{D8:}\,\, & \frac{1}{\Lambda^2} \bar\chi \gamma_\mu\gamma^5 \chi\;\bar q \gamma_\mu\gamma^5 q
\end{aligned}
\end{equation}
where the nomenclature was first adopted in~\cite{Goodman:2010ku}.

Let us briefly point out, and we will come back to this in Sec.\ref{sec:simpmodsdiscussion}, that the scalar and pseudo-scalar models of Eqs.~\ref{eq:S},~\ref{eq:P} are not gauge invariant.
This may lead to spurious results in processes where a $W/Z$ boson is emitted.
Moreover, in the axial vector model perturbative unitarity is violated in a large portion of parameter space.
We will return to these issues in section~\ref{sec:simpmodsdiscussion}.

\begin{figure}
\centering
\includegraphics[scale=1]{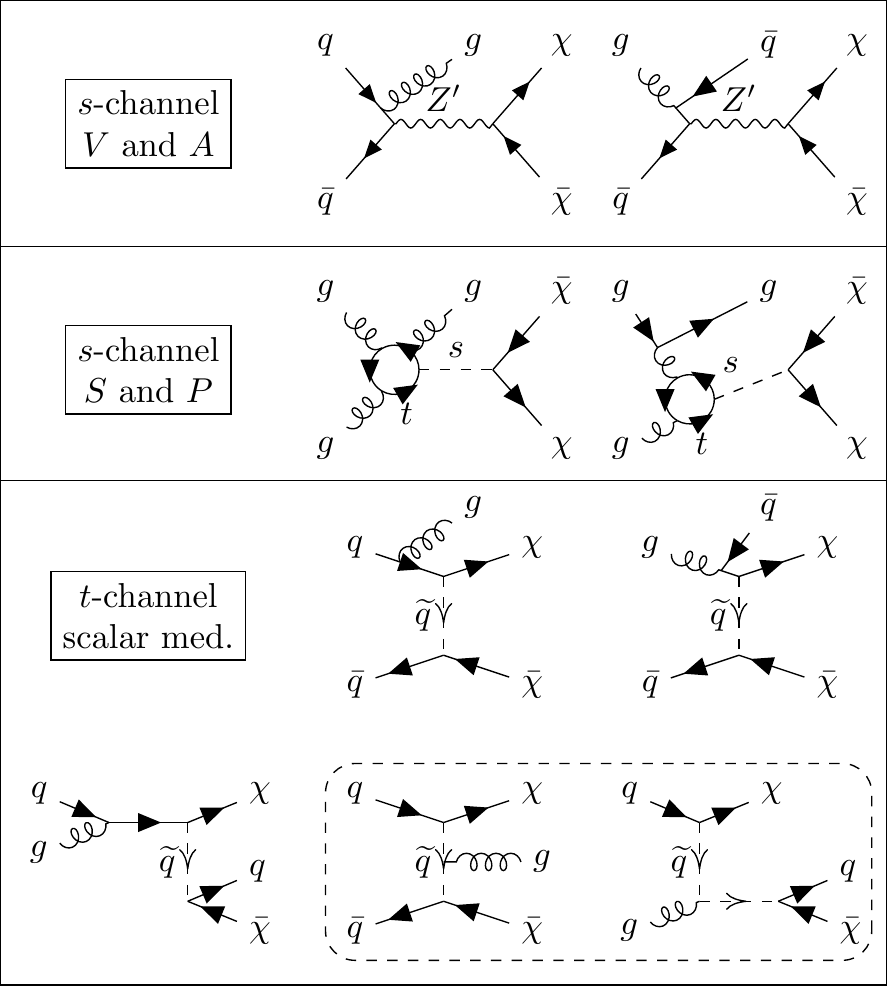}
\caption{Feynman diagrams for the production of a DM pair in association with a quark or a gluon, leading to a mono-jet signature. Additional diagrams obtained by exchanging a fermion and a anti-fermion, as well as the ones obtained by permuting the gluon vertices in the loop in the (pseudo-)scalar case, are neglected. The diagrams enclosed in the dashed box in the $t$-channel model are suppressed in the EFT limit.}
\label{fig: diagrams monojet}
\end{figure}

Consistently with the MFV hypothesis, we force the couplings to be diagonal: $g_q^{ij}=g_q^i\delta^{ij}$.
Moreover, we assume them to be flavour-blind in the (axial-)vector case and proportional to the SM Yukawa in the (pseudo-)scalar ones:
\begin{equation}
g_d^i=g_u^i\equiv g_q \,,
\quad 
y_q^{ij} \equiv y \frac{m_i}{v} \delta^{ij}
\quad {\rm for} \quad i=1,2,3\, .
\end{equation}
In this way the spin-0 models have an enhanced coupling to the third generation's quarks, which makes the phenomenology quite different from the spin-1 models, both because of the different production mechanism (gluon fusion with a top loop instead of $q\bar q$ annihilation) and because of the possibility of constraining such models with searches in $b/t$ channels.
As an example, Fig.~\ref{fig: diagrams monojet} shows the Feynman diagrams involved in the calculation of the mono-jet cross section.

In addition to the parameters of the Lagrangian, in the calculation of scattering amplitudes one must also include the decay width $\Gamma$ of the mediator, which can be thought as a free parameter that encodes the unknown decay probability to other particles belonging to the dark sector. In this case, in the computation of the cross sections the couplings constants factor out and they affect only the normalization of the cross section through their product $g_q g_\chi$, while the spectra depend only on the masses $\mmed, m_\chi$.
The problem with this approach is that, typically for large values of $\mmed$, the benchmark value of $\Gamma$ becomes smaller than the sum of the partial widths for decays into DM and quarks, which makes the choice unphysical~\cite{Busoni:2014gta}.
To avoid the problem it is usually assumed that the mediator can not decay into particles other than the SM ones and, depending on its mass, the DM, and the width is computed accordingly as
\begin{equation}
\Gamma = \Gamma_\chi + \sum_f \Gamma_f + \Gamma_{gg}
\end{equation}
This is usually referred to as the ``minimal width assumption''.
Even if this choice eliminates one parameter, its drawback is that now the cross sections depend non-trivially on the couplings:
\begin{equation}
\sigma \propto \frac{g_\chi^2 g_q^2}{(s-\mmed^2)^2+\mmed^2\Gamma^2}
\end{equation}
where $\Gamma$ at the denominator depends on $g_\chi, g_q$. Nevertheless, in most cases the dependence on the couplings is less important than the one on the masses, and so it is a good choice to fix the couplings and let the mass vary, thus presenting results as exclusion plots in the plane $m_\chi$ \textit{vs.} $\mmed$.
Following the recommendations of the LHC Dark Matter Working Group, useful benchmarks for the vector (V) and axial vector (A) models are~\cite{Boveia:2016mrp, Albert:2017onk}:
\begin{equation}
V:\,\left\{
\begin{array}{l}
\gX=1, g_q=0.25, g_\ell=0 \\
\gX=1, g_q=0.1, g_\ell=0.01 \\
\end{array}\right.
\quad
A:\,\left\{
\begin{array}{l}
\gX=1, g_q=0.25, g_\ell=0 \\
\gX=1, g_q=0.1, g_\ell=0.1 \\
\end{array}\right.
\end{equation}

\medskip

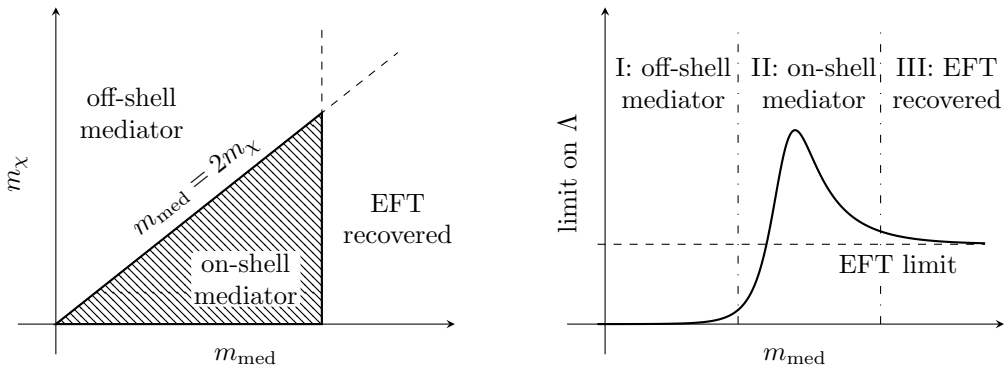
\begin{figure}
\centering
\begin{tikzpicture}
\begin{axis}[
domain=-.02:1,
xmin=-.1, xmax=1.05,
ymin=-.1, ymax=1.05,
x=5cm,
y=4cm,
samples=400,
axis y line=middle,
axis x line=middle,
x label style={at={(axis cs:.5,-.05)},anchor=north},
y label style={at={(axis cs:-.05,.5)},anchor=south, rotate=90},
xlabel={$\mmed$},
ylabel={$m_\chi$},
ticks=none
]
%	\addplot+[mark=none, dashed] {x};
	\addplot[dashed] coordinates {(.7,0)(.7,1)};
	\addplot[dashed] coordinates {(0,0)(.9,.9)}
		node[pos=.45,sloped,above,align=center] {$\mmed=2m_\chi$};
	\draw [thick, pattern=north west lines, pattern color=black] (axis cs:0,0) -- (axis cs:.7,0) -- (axis cs:.7,.7) -- cycle;
	\node[align=center] at (axis cs: .9,.35) {EFT\\ recovered};
	\node[align=center, fill=white, rounded corners=2pt,inner sep=1pt] at (axis cs: .5,.15) {on-shell\\ mediator};
	\node[align=center] at (axis cs: .2,.7) {off-shell\\ mediator};
    \end{axis}
\end{tikzpicture}
\hspace{1cm}
\begin{tikzpicture}
\begin{axis}[
domain=-.02:1,
xmin=-.1, xmax=1.05,
ymin=-.1, ymax=1.05,
x=5cm,
y=4cm,
samples=400,
axis y line=middle,
axis x line=middle,
x label style={at={(axis cs:.5,-.05)},anchor=north},
y label style={at={(axis cs:-.05,.5)},anchor=south, rotate=90},
xlabel={$\mmed$},
ylabel={limit on $\Lambda$},
ticks=none
]
    \addplot+[mark=none,black,thick] {0.45*pow(10,8*(x+1))/((1000^2-pow(10,4*(x+1)))^2 + 0.7*pow(10,8*(x+1)))};
    \addplot[dashed, black] {0.45*0.588};
	\draw[loosely dash dot] (axis cs:.35,0) -- (axis cs:.35,1);
	\draw[loosely dash dot] (axis cs:.725,0) -- (axis cs:.725,1);
	\node[align=center] at (axis cs: .9,.8) {III: EFT\\ recovered};
	\node[align=center, fill=white, rounded corners=2pt,inner sep=1pt] at (axis cs: .55,.8) {II: on-shell\\ mediator};
	\node[align=center, fill=white, rounded corners=2pt,inner sep=1pt] at (axis cs: .18,.8) {I: off-shell\\ mediator};
	\node[anchor = north east, fill=white, rounded corners=2pt,inner sep=3pt] at (axis cs: 0.95,0.26) {EFT limit};
    \end{axis}
\end{tikzpicture}
\caption{\textit{Left:} Sketch of the limits obtained from mono-X analysis in the $\mmed - m_\chi$ plane (\textit{left}, adapted from~\cite{Kahlhoefer:2017dnp}) and in the $\mmed - \Lambda$ plane (\textit{right}, adapted from~\cite{Buchmueller:2013dya}).}
\label{fig: limits monojet}
\end{figure}

The exclusion lines that LHC draws have typically a simple structure. In MET+X searches in which DM is pair produced from the mediator and recoils against a SM particle (a photon, a hadronic jet or other) that is necessary to tag the event, the best sensitivity is obtained for $\mmed > 2m_\chi$, where DM can be produced on resonance and the cross section is consequently enhanced. On the other hand, for $\mmed < 2m_\chi$, the cross section is suppressed.\footnote{
In the mono-jet channel, for $\mmed \lesssim 2 \mX$, the LHC at $14\TeV$ with $300\ifb$ is sensitive to $\mathcal{O}(1)$ couplings only for $\mX\lesssim\mathcal{O}(100\GeV)$, while for $\mX\sim1\TeV$ it is sensitive only to couplings of order $\gX\cdot g_q \gtrsim 10$~\cite{Abdallah:2014hon}.}
Finally, for large $\mmed$ the EFT limit is recovered, but then again the constraining power is suppressed by the large $\mmed^4$. Monojet limits in this region extend up to around $1.5-2\TeV$, depending on the search, on the choice of vector or axial-vector mediator and to the values of the couplings (see \eg Figs. 3 and 4 of~\cite{Choudhury:2015lha} for the dependence of the $\mmed$ limit on the values $g_q,g_\chi$).
This give rise to a typical triangular shape in the exclusion plots sketched in Fig.~\ref{fig: limits monojet} (see \eg~\cite{Sirunyan:2017jix} for a very recent example of such an exclusion plot in the mono-$Z/W$ channel).

Sometimes, it proves very useful to show the constraints on $s$-channel simplified models in the  plane $\mmed - \Lambda$ for fixed $\mX$, where $\Lambda = \mmed/\sqrt{g_q g_\chi}$ is defined as the contact interaction scale. In this plane, the bounds have the typical shape shown in the right panel of Fig.~\ref{fig: limits monojet}. For light mediator (Region I), DM production proceeds off-shell, and the cross section is suppressed (compared to the corresponding EFT result) by $(\mmed^4/s^2)$, where $\sqrt{s}$ can be estimated as $\min(\textrm{MET}^2, \mX^2)$.
In Region II, the mediator is produced on-shell, and the cross section is enhanced. In this region the limit depends on the choice of the width $\Lambda$, as the cross section scales as $g_\chi^2 g_q^2 / (\mmed^2\Gamma^2)$.
Finally, in Region III the EFT limit is recovered.

\medskip

Missing transverse energy searches are not the only handle that we have on simplified models. Searches for the mediator, for example in the resonant di-jet channel, can lead to more stringent bounds in $\mmed$ for the same value of the coupling to quarks $g_q$.

Constraints on $s$-channel simplified models have been obtained by numerous groups, with a particular attention to the case of a (axial-)vector mediator, due to the problematic nature of the (pseudo-)scalar models of Eq.~\ref{eq:S}, \ref{eq:P} (see the discussion in section~\ref{sec:simpmodsdiscussion}).
Mono-jet constraints were discussed in~\cite{Lebedev:2014bba, Buchmueller:2014yoa, Harris:2014hga, Buckley:2014fba, Xiang:2015lfa, Harris:2015kda}.
A thorough comparison of mono-jet searches to dijet searches, direct detection limits, dark matter overproduction in the early universe and constraints from perturbative unitarity is performed in~\cite{Fairbairn:2014aqa, Chala:2015ama, Choudhury:2015lha, Duerr:2016tmh, Fairbairn:2016iuf}.
In~\cite{Jacques:2015zha,Brennan:2016xjh} the problem of deriving limits for arbitrary values of the coupling constants starting from the benchmark ones is addressed.
The case for a light DM (thus evading constraints from direct searches) is analysed in~\cite{Gondolo:2011eq, An:2012va, An:2012ue}.

A very interesting phenomenology arises in the case where the couplings to third generation quarks is larger than the couplings to the first two, as it is the case in the spin-0 models with MFV introduced above.
Strong constraints on these models come from searches for one or two $b$-tagged jet~+~MET and $t\bar t$~+~MET (see \cite{Lin:2013sca} for an early proposal within the EFT framework and \cite{Buckley:2014fba, Abercrombie:2015wmb, Haisch:2015ioa, Arina:2016cqj, Haisch:2016gry, CMS:2016mxc, CMS:2016jxd} for a discussion in terms of simplified models).
Summarizing, for light DM ($\mX=1\GeV$) current bounds obtained in the $t\bar t$~+~MET channel can exclude couplings $g\gtrsim 1$ up to $\mmed\lesssim100\GeV$~\cite{CMS:2016mxc, CMS:2016jxd}, while this value can  decrease to $\approx 0.5$ at the end of the planned LHC runs~\cite{Haisch:2016gry}.
These values are similar in magnitude to the ones obtained by a mono-jet analysis (see \eg~\cite{Arina:2016cqj}), which are much weaker than in the (axial-)vector case due to the assumed SM-Yukawa-like structure, which suppresses the coupling to light quarks. Still, LHC searches can provide the most stringent limit in some region of parameter space, complementing those coming from direct detection and from the relic abundance constraint, and proving once more the importance of the complementarity of different probes~\cite{Arina:2016cqj}.

\subsection{$t$-channel mediators}

Another interesting possibility is that of a coloured fermionic mediator with an interaction vertex between quarks and the WIMP resulting in a $t$-channel exchange, as with squark in supersymmetric models:
\begin{equation}\label{eq:t-ch lagrangian}
\Lagr = \Lagr_\mathrm{SM} + \sum_{i} \left(g_L^i \bar{Q}^i_L \widetilde{Q}_L^i  +  g_u^i \bar{u}^i_R \tilde{u}^i_R + g_d^i \bar{d}^i_R \tilde{d}^i_R \right)\chi + \text{mass terms} +c.c. \,,
\end{equation}
where $Q_{L}^i, u_R^i, d_R^i$ are the usual SM quarks, $\widetilde{Q}_L^i, \widetilde{u}_R^i, \widetilde{d}_R^i $ correspond to the respective scalar mediator (the squarks), and $i$ represents a flavour index.
Unlike the usual case in supersymmetry, here the WIMP $\chi$ can be taken to be either Dirac or Majorana fermion.
This model is extensively analysed in~\cite{Bell:2012rg, Chang:2013oia, An:2013xka, Bai:2013iqa, DiFranzo:2013vra, Papucci:2014iwa, Garny:2014waa, Garny:2015wea}.
While in the model above the flavour index is carried by the mediators, it could be the case that this index is assigned to the DM itself. This is the so-called ``flavoured DM'' scenario~\cite{Agrawal:2014una, Batell:2013zwa}

As it can be easily understood, also in the case of $t$-channel mediators the phenomenology depends on the relative values of the couplings and of the masses involved.
The minimal flavour violation hypothesis forces the couplings $g^i$ and the masses of the scalars $m_i$ to be equal. This assumption can be relaxed for the third generation. Particularly interesting is the case in which the coupling to the third generation is enhanced, and strong constraints come from searches for $b,t$ quarks in the final state~\cite{Andrea:2011ws, Kamenik:2011nb, Kumar:2013hfa, Batell:2013zwa, Agram:2013wda, Abdallah:2014hon, Kilic:2015vka, Garny:2017rxs, Garny:2018icg}.
A distinct phenomenology arises in the case of small couplings, in which the relic density is obtained by a out-of-equilibrium freeze-out mechanism and exotic collider signatures such as disappearing tracks and displaced vertices~\cite{Garny:2017rxs, Garny:2018icg}.

Two interesting features of this model are worth listing, that makes it qualitatively different from its low energy EFT limit.
Firstly, being the squarks coloured, gluons may be emitted not only as initial state radiation but also from the mediator itself.
This process is suppressed in the EFT limit by two powers of $\Mmed$, and this make a large qualitative difference in the kinematic distribution within the simplified model and the corresponding operator.
Secondly, when the mediator is light enough, its pair production becomes kinematically accessible, and an event like $p\,p \to \widetilde{q}\,\widetilde{q} \to q\chi \,q\chi$ leads to a di-jet + MET signature (or in general jets + MET, when additional jet radiation is taken into account).
Interestingly, this signature with two high-$\pT$ jets leads to constraints stronger then the mono-jet one on a large portion of parameter space, except for the compressed region $|\mmed-m_\chi|\ll\mmed$~\cite{Papucci:2014iwa}.

An interesting phenomenology arises in the case in which the $t$-channel mediators couple DM to both quarks and leptons, as in~\cite{Altmannshofer:2014cla, Capdevilla:2017doz}. Radiative corrections in this model strongly alter the spectrum of the Drell-Yan process $q \bar q \to \ell^+ \ell^-$. With this signal, one can probe ``compressed regions'' that are difficult to probe in direct searches for mediators like jets + MET searches. Moreover, these features can be used to to make qualitative statements about dark matter’s self-conjugation, mass, spin, and chirality of interactions.

\subsection{Other models}

In addition to the models listed above, many interesting ones may be constructed that cannot be addressed here. Those include spin-2 mediators~\cite{Kraml:2017atm}, t-channel fermionic mediators, fermiophobic scalar mediators~\cite{Englert:2016joy}, gluphylic mediator models~\cite{Godbole:2015gma, Godbole:2016mzr, Ducu:2015fda}, models with SM portals (Higgs or $Z$)~\cite{Arcadi:2014lta}, models of scalar DM~\cite{Athron:2017kgt}, vector DM~\cite{Arcadi:2016qoz, Arcadi:2017jqd}, Higgs portal models with DM of diverse spin number~\cite{He:2016mls, Chang:2017gla, Chang:2017dvm} and others. A recent and comprehensive review is given in~\cite{Arcadi:2017kky}.
Interesting cosmological features of a model in which DM couples predominantly to the top quark are explored in~\cite{Delgado:2016umt}. In this case, it is possible to obtain the correct abundance with annihilations of DM particles to heavier states, at the tail of the velocity distribution.
%

%%%%%%%%%%%%%%%%%%%%%%%%%%%%%%%%%
\section{Less simplified models}
\label{sec:simpmodsdiscussion}
%%%%%%%%%%%%%%%%%%%%%%%%%%%%%%%%%

\subsection{A critical look}

The simplified models that we discussed so far can be viewed as an improvement of effective operators, where the effective scale $\Lambda^4$ is replaced by a propagator's denominator $(p^2-M^2)^2+\Gamma^2 M^2$ in order to avoid energy limitations and exploit resonant enhancement in the production cross section. This is one step above in a bottom-up approach.
Nevertheless, the models described above suffer from other limitations, and are not fully self consistent, as we are going to illustrate in a moment.

The reason why the theoretical consistency of the simplified models is important is twofold. On the one hand, violation of perturbative unitarity can lead to spuriously large predictions (\eg in the process of $W$ emission), leading to artificially strong bounds on the parameter space of the model.
On the other hand, thinking of a full UV completion from which simplified models may descend, theoretical consistency is a necessary requirement at the level of the full theory. One could argue that this may not be the case for the simplified model, since other fields belonging to the full theory may restore the desired consistency.
While this is generically true, these additional fields may add interesting ingredients to the phenomenology of the model, as they may produce new final states at the LHC and modify the annihilation cross section that enters the relic abundance calculation and the Indirect Detection fluxes, and their inclusion is therefore mandatory.

\subsubsection{Gauge non-invariance of the the (pseudo-)scalar model}
\label{sec: scalar gauge invariance}
The scalar and pseudo-scalar models of Eqs.~\ref{eq:S}, \ref{eq:P} are manifestly not gauge invariant. The problem comes with the Yukawas of the scalar mediator: assuming the DM is a singlet under the SM group, the invariance of the term $S\bar\chi\chi$ forces $S$ to be a singlet as well, while for the Yukawas with the SM quarks $S\bar q q$ and $P\bar q\gamma_5 q$ the mediator should transform as a doublet. Clearly such a model can exist only as a consequence of EW symmetry breaking, and the DM-SM interaction has to be suppressed at low energy by some power of $v_\textrm{\small EW}/\mmed$~\cite{Bell:2015sza, Baek:2015lna}. Clearly, a UV completion of the model is necessary.

As we will detail below, this issue can be fixed if the mediator is assumed to couple at tree level \emph{only} to the DM and to the Higgs through the gauge invariant portal terms $S |H|^2$, $S^2|H|^2$, \etc that gives rise to a non zero mixing angle after EW symmetry breaking~\cite{Baek:2011aa, Baek:2012uj, Baek:2015lna, Bell:2016ekl}. This construction can be replicated for a pseudo-scalar, with the advantage of avoiding strong Direct Detection bounds, but at the price of introducing a new source of CP violation~\cite{Ghorbani:2014qpa, Ghorbani:2016edw, Baek:2017vzd}.

In~\cite{Baek:2015lna}, the model of Eq.~\ref{eq:S} in compared to three possible consistent variations of it: first, the one in which the coupling of the scalar mediator to SM quarks is suppressed by $v_\textrm{\small EW}/\mmed$; second, the model in which the mediator is replaced by the Higgs itself and the coupling to the DM particle is suppressed by the same factor; finally, the Higgs portal model including the $S-h$ mixing mentioned above.
Naively, one could thing that the $S$-mediator and the $h$ mediator models can provide good approximations of the Higgs portal model. This is true only for large $S$ mass ($m_S\gtrsim 1\TeV$ for $\mX=50\GeV$, or $m_S\gtrsim 5\TeV$ for $\mX=400\GeV$), in which case the Higgs portal resembles the model with the Higgs as a mediator, because the heavy scalar $S$ can be integrated out. On the other hand, for lower masses the behaviour of the Higgs portal model descends from the interplay of the two mediating particles $h$ and $S$, and the limits can be both stronger or weaker than the ones obtained in the two single-mediator models.
In the case $\mX > m_h/2$, where the mixing angle is not constrained by the Higgs-to-invisibles branching ratio, the DM production cross section is dominated by $S$ exchange, and in the heavy $S$ limit a bound $M_\star\gtrsim 20\GeV$ can be imposed, where
\begin{equation}
\frac{1}{M_\star^3}\approx \frac{\lambda \sin\alpha\cos\alpha}{v_H m_h^2} \,,
\end{equation}
$\alpha$ being the mixing angle and $\lambda$ the coupling of the scalar $S$ to the DM particle~\cite{Baek:2015lna}.

A model of this kind naturally replicates the SM Yukawa-like structure of Eqs.~\ref{eq:S}, \ref{eq:P}, and displays promising experimental signatures (mono-jet, heavy quarks, mono-V, see~\cite{Kahlhoefer:2017dnp} and references therein). Nevertheless, the mixing angle $\epsilon$ is strongly constrained by Higgs physics measurements (in particular the decay rate into invisible particles and the Higgs signal strength, that are respectively enhanced and reduced by a non zero mixing angle), and it does not add anything new to the LHC models' toolbox. An interesting option is to extend it to a two-Higgs doublets model (2HDM) with the addition of a singlet scalar, evading all such constraints~\cite{Goncalves:2016iyg, Bell:2016ekl}.

\subsubsection{Gauge non-invariance and violation of perturbative unitarity}
The vector and axial-vector simplified models of Eqs.~\ref{eq:V}, \ref{eq:A} are not in general invariant under the full SM gauge group ${\rm SU}(3)_c\times{\rm SU}(2)_L\times{\rm U}(1)_Y$ but only under the unbroken subgroup ${\rm SU}(3)_c\times{\rm U}(1)_{\rm e.m.}$. In particular, if the couplings to up and down quarks are different the mediator does not couple to the left handed quark doublet but to its two components separately, thus breaking gauge invariance.
	Similarly, the $t$-channel model of Eq.~\ref{eq:t-ch lagrangian} is not gauge invariant unless the scalar mediator $\widetilde{Q}_L^i$ is charged and transforms as $({\bf 2},-1/2)$ under ${\rm SU}(2)_L\times{\rm U}(1)_Y$.
	Violation of the electroweak gauge symmetry can lead to spuriously enhanced cross section for DM production with the initial state radiation of a $W$ boson~\cite{Bell:2015sza,Bell:2015rdw}. This problem does not only affect the mono-$W$ searches: the $W$ can indeed decay hadronically, enhancing the signal in the mono-jet search. For example, in the case of a vector mediator with opposite sign couplings to up and down quarks, this process dominates the mono-jet cross section for $\met>400\GeV$~\cite{Haisch:2016usn}. Similar issues should be present when considering $Z$ or $\gamma$ emission.
	In passing, this example shows that constraints descending from the internal consistency of the model can not be neglected even when restricting to a particular MET search such as the mono-jet one, that at first sight looks safe.
	Referring to the case of a vector mediator, different couplings of the up and down quarks can be made compatible with perturbative unitarity if an appropriate vertex $WWZ'$ is added (where $Z'$ is the new vector mediator), in similarity to what happens for the $Z$ boson in the SM. In the t-channel model, instead, perturbative unitarity is restored if $W$ emission from the charged mediator's line is included in the calculation.

\subsubsection{Violation of perturbative unitarity with a s-channel axial-vector mediator}

In the axial vector model~\ref{eq:A} the coupling to the longitudinal mode of the mediator is enhanced for heavy fermions by the ratio $m_f/m_A$. In particular, considering the elastic scattering of fermions (both SM fermions or DM) the perturbative unitarity bound on this model reads $m_f \lesssim \mZp/(\sqrt2 g_f^A)$~\cite{Kahlhoefer:2015bea}, where $f$ may stand for both a SM fermion or the DM particle.
Even if such a bound is satisfied, perturbative unitarity is still violated in the process of 2 fermions annihilation into $Z'Z'$, which is important for the calculation of the relic density and for indirect detection.
In order to restore unitarity some additional ingredient has to be invoked. In particular, what violates unitarity is the longitudinal mode of the $Z'$ boson, therefore the addition to the model of a scalar particle, invariant under the SM gauge group, that give rise to its mass via Higgs mechanism serves the purpose.
In this case, the condition on the mass of the $Z'$ would be~\cite{Kahlhoefer:2015bea}
\begin{equation}
\sqrt{\pi}\frac{\mZp}{g_{\rm DM}^A} \geq \max [m_s, \sqrt{2}m_{\rm DM}] \,,
\end{equation}
where $m_s$ is the mass of the new scalar. At this point, it is clear that such an issue is not present in the vector model, because the mass of the mediator in that case can be obtained via a Stueckelberg mechanism without the need of additional particles (see~\cite{Bell:2016uhg} for further discussion).

\subsubsection{Invariance of the SM Yukawas}\label{sec: Higgs charged}

Again referring to the axial-vector model of Eq.~\ref{eq:A}, if this is thought as a gauge extension of the SM, then the SM fermions must be charged under the new gauge symmetry (we restrict for simplicity to a $\textrm{U}(1)'$ theory, often referred to as a ``$Z'$ model'').
	Therefore, the Yukawa terms $H \bar Q_L d_R$ are gauge invariant only if the Higgs is charged as well, with $q_H = q_{qL}-q_{uR} = q_{dR}-q_{qL} = q_{eR}-q_{\ell L}$ (which is zero only if the SM fermions are vector-like under the new symmetry, as in the $B-L$ case)~\cite{Kahlhoefer:2015bea}.
	From this relation one sees that leptons have to be charged in this model, thus resulting in strong constraints from di-lepton searches.
	Another important consequence is that, after electroweak-symmetry breaking, the SM $Z$ and the $Z'$ have a non zero mixing angle, and a tree level $hZZ'$ vertex appears, with important phenomenological consequences and a complicate interplay between the two effects~\cite{Jacques:2016dqz,Cui:2017juz}.
%	Moreover, in this 

%
\subsubsection{Cancellation of gauge anomalies}
If the interaction of DM with SM fermions is due to an extended gauge symmetry, in order for the theory to be consistent at the quantum level the charge assignment under the new gauge group cannot be generic. If all the fermions of the dark sector are uncharged under ${\rm SU}(3)_c\times{\rm SU}(2)_L\times{\rm U}(1)_Y$, then the SM ones must have charges chosen in such a way to cancel the mixed anomalies of the dark gauge group with the SM. It can be shown that for a $\textrm{U}(1)'$ theory this forces the charges to be a linear combination of the SM hypercharge $Y$ and of $B-L$~\cite{Weinberg:1996kr}. This implies that the mediator must couple to leptons, leading to tight constraints from resonance searches in the dilepton channel (see \eg~\cite{Jacques:2016dqz}).
	
	Alternatively, additional heavy fermions, charged under the SM, may be added to the model. The mass of these fermions cannot be arbitrarily large: in order to cancel the anomalies, they must be chiral at least under the dark gauge group, and their mass is given by the vev of the dark Higgs.
	Fortunately, imposing the invariance of the SM Yukawa terms as discussed above in Sec.~\ref{sec: Higgs charged}, it turns out that the gluon-gluon-$Z'$ anomaly automatically cancels, and the new states need not to be coloured, reducing their impact on LHC searches~\cite{Kahlhoefer:2015bea}.
	On the other hand, they will enter the calculation of loop induced processes such as two photons decays, that are not calculable in an anomalous theory, and are relevant for indirect searches~\cite{Duerr:2015vna, Jacques:2016dqz}.\footnote{Anomalous models can be studied within a specific EFT framework, in which the required additional heavy fermions are integrated out, resulting in a set of effective Chern-Simons terms that must be added to the theory~\cite{Ismail:2017ulg}.}

\subsection{The $\mathbf{U(1)'}$ model}
This model is one of the simplest possible extensions of the SM, in which the gauge group is enlarged by an additional $\mathrm{U}(1)'$, spontaneously broken by the vev of a scalar field $s$, singlet under the SM, that gives mass to the dark gauge boson $Z'$.
As mentioned above, interesting features come from this construction. First, the invariance of the SM Yukawas force the Higgs to be charged under $\mathrm{U}(1)'$ whenever the charges of the left- and right-handed SM fermions differ from one another.
Second, and consequently, the Higgs kinetic term includes interactions with the $Z'$ boson, that induce after EW symmetry breaking a $ZZ'h$ that may have an impact on indirect detection~\cite{Jacques:2016dqz}.
Moreover, the dark Higgs $s$ enriches the phenomenology, both at the LHC and in the calculation of annihilation rates.
Finally, in consistent models that implement gauge anomaly cancellation the SM leptons must typically be charged and couple to the $Z'$, which is therefore constrained by resonant dilepton searches.%
\footnote{The coupling to leptons can be avoided by adding new fermions with non trivial transformations under the SM gauge group such that all anomalies cancel~\cite{Ellis:2017tkh}, or in models where the DM carries a non-zero baryon number~\cite{Duerr:2013dza, Duerr:2013lka, Perez:2014qfa, Ohmer:2015lxa}.}
This constraint may be evaded if the $Z'$ is lighter than the SM $Z$, thus escaping resonant searches~\cite{Alves:2016cqf}.

A number of studies have addressed DM $Z'$ models with respect to LHC, direct and indirect searches, as well as its cosmological implications~\cite{Alves:2013tqa, Arcadi:2013qia, Alves:2015pea, Ghorbani:2015baa, Alves:2015mua, Duerr:2016tmh, Arcadi:2017hfi}.
If DM is a Dirac fermion, its non-relativistic scattering off nuclei is spin-independent, and direct detection constrain the DM mass to be larger than $\sim1\TeV$~\cite{Alves:2015mua}. Such a strong constraint is lifted if the DM is a Majorana fermion, since the vector bilinear $\bar\chi\gamma_\mu\chi$ vanishes exactly and only the axial-vector one is left.
In models in which the mediator couples to leptons with a coupling similar in magnitude to the one to quarks (as required by anomaly cancellation) resonant dilepton searches forces the $Z'$ to be heavier than $\sim 3-4 \TeV$, depending on the couplings and on the structure of fermionic charges~\cite{Aaboud:2017buh}. Dijet searches can exclude $\mZp\lesssim 2.5\TeV$ for a coupling $g_q\approx0.25$, see~\cite{Aaboud:2017yvp}.
Monojet searches are typically weaker than the dijet ones, reaching $\mZp\lesssim 1.5\TeV$ for $g_q=0.25, g_\chi =1$ and for $\mX<\mZp/2$~\cite{ATLAS:2017dnw}.

The dark Higgs becomes relevant when the annihilation rate is concerned. Indeed, the $s Z'$ annihilation channel proceeds in $s$-wave and, therefore, it dominates the cross section for indirect detection and for the relic abundance calculation~\cite{Bell:2016fqf}, together with the $ss$, $Z'Z'$ and $Zh$ channels~\cite{Bell:2016fqf, Duerr:2017uap}. In particular, the additional channels and operators increase the DM annihilation rate, thus reducing the relic abundance and alleviating constraints from DM overproduction in the early universe.
Moreover, mono-dark-Higgs signals can be looked for at the LHC, with the typical signature being DM produced in association with a scalar resonance $s$ decaying to a highly boosted $b\bar b$ pair~\cite{Duerr:2017uap}. In this case, the expected LHC sensitivity extends up to  $\mZp\lesssim 3.5\TeV$, $\mX\lesssim 600\GeV$ for $m_s=50\GeV$, or higher for heavier $s$.
In general, such a model can be viewed as a model with two mediators, that in some limit may reduce to a spin-1 or to a spin-0 mediator~\cite{Duerr:2016tmh}.

\subsection{Two Higgs doublets (plus one singlet) models and DM}

Models in which the (pseudo-)scalar mediator that couples to DM obtains its coupling to SM quarks from mixing with a second Higgs doublet have received a significant attention recently. With respect to the scenario in which the singlet mixes directly with the SM Higgs discussed in Sec.~\ref{sec: scalar gauge invariance}, in this model the Higgs branching ratios and signal strength are not modified.
Moreover, both a scalar and a pseudoscalar mediator can be accomodated without adding new sources of CP violation. Indeed, in the presence of a second doublet, there are $8$ spin-0 fields, three of which get `eaten' by the $Z,W^\pm$ after symmetry breaking, thus leaving one charged scalar field, two neutral ones and one neutral pseudoscalar, which the dark mediator can mix with.
The pseudoscalar case is particularly interesting in view of the fact that its low energy effective vertex $\bar q i \gamma_5 q \bar\chi i\gamma_5\chi$ leads to the both spin- and momentum-suppressed non relativistic interaction $(\vec s_\chi\cdot\vec q)(\vec s_N\cdot\vec q)$~\cite{Fan:2010gt, Fitzpatrick:2012ix}, on which constrains from Direct Detection are poor.

A thorough discussion of 2HDM (not related to DM) is given in~\cite{Branco:2011iw}.
The model of interest here, where the 2HDM is complemented with an additional pseudo-scalar mediator and with a DM fermion, is described in details in~\cite{Bauer:2017ota}.
In general, the model counts many free parameters, many of which can be fixed by requiring that Higgs and precision EW tests are not spoiled, and by the requirement of the stability of the scalar potential. In particular, one of the two Higgs states is assumed to have SM-like couplings, while the second doublet couples to SM vectors only at loop level: this is the so-called \emph{alignment/decoupling limit}. Moreover, the neutral scalar, pseudoscalar, and the charged components of the second Higgs doublet are assumed to have the same mass. A typical choice is then to fix the DM Yukawa coupling to $1$, and the DM mass to a benchmark value of $10\GeV$. The model then consists of 4 free parameters, which are typically chosen to be the mass of the pseudoscalars $m_a, m_A$, the ratio of the VEVs $\tan\beta = v_1/v_2$ and the $a-A$ mixing angle $\theta$.

In the context of DM, many search channels have been used to constrain this class of models. In particular, searches for mono-$\mathrm{jet}/\gamma/Z/W/H$, non-resonant dijets, single top as well as $t\bar t$, $b\bar b$, $b\bar b Z$, all in association with missing transverse energy, have been explored~\cite{No:2015xqa, Goncalves:2016iyg, Haisch:2016gry, Angelescu:2016mhl, Bell:2016ekl, Baek:2017vzd, Bauer:2017ota, Banerjee:2017wxi, Tunney:2017yfp, Bell:2017rgi, Arcadi:2017wqi, Pani:2017qyd}. As usual, constraints come also from visible channels (\ie where no invisible DM particle is produced), as well as indirect detection and cosmology. Finally, all the usual concerns about 2HDM coming from flavour physics, EW precision tests, invisible Higgs decays, vacuum stability and perturbative unitarity apply too.

The Higgs' width to invisibles constrains $m_a\gtrsim 100\GeV$, while flavour constraints forces $\tan\beta \gtrsim 1$. The most powerful searches are then the mono-Z and the mono-H ones, that can exclude $\tan\beta$ up to $2$ for values of $m_a$ up to $200-350\GeV$, depending on the values of $m_A$ and $\sin\theta$. For further details we refer the reader to~\cite{Bauer:2017ota}.

Interestingly, since in such a model the coupling of DM to the heavy quarks is naturally enhanced, the Galactic Centre excess could be explained by a model which is testable at the LHC~\cite{Ipek:2014gua, Tunney:2017yfp}.

Let us just mention another related possibility, which is the one of a second inert Higgs doublet, which does not couple to SM fermions except for its mixing with the Higgs. The lightest component of this doublet is a perfect candidate for scalar DM~\cite{LopezHonorez:2006gr}.

\section{Conclusions}
\label{sec:conclusions}

The question of which DM models should be adopted in defining new search strategies and in presenting experimental results is a pressing one, primarily for LHC searches.
Simplified DM models are a possible answer to this question, living in between the effective operators approach (with a limited applicability at the LHC) and the realm of well motivated BSM theories.

From a bottom-up viewpoint, the idea of simplified models is to expand the effective operators including mediator particles in the description, thus avoiding the energy limitations of the EFT approach and adding a richer phenomenology, new search channels, {\it etc}. In a top-down framework, instead, simplified models can be seen as a way to simplify the phenomenology of complex new physics models in such a way to restrict to the phenomena related to DM.

In order not to deal with unphysical results, the vanilla simplified models have to be supplied with additional constraints, couplings and states, in a kind of second order improvement. The typical consequence is that the strongest LHC constraints on the dark sector come from many possible observables other than DM production processes (as the mono-X searches) and di-jet searches (\eg di-lepton resonances, mixing with $Z$ boson and electroweak precision tests, Higgs width to invisibles, perturbative unitarity and so on).
This comes with no surprise, since the high energy reach of the LHC consents to explore a large variety of phenomena above the weak scale, without restricting to the lightest stable state of this new physics sector. This is quite the opposite with respect to what happens with direct and indirect searches, which are intrinsically limited to constrain the properties of the DM particle.

Theoretically consistent simplified models tend to loose part of their generality, and to mimic richer BSM theories. For example, models containing a vector mediator and a dark Higgs may descend from gauged $\mathrm{U}(1)'$ constructions, while models featuring two Higgs doublets and a (pseudo-)scalar singlet resemble the Higgs sector of the NMSSM.

Simplified models cannot (or only partially) be viewed as an exhaustive toolbox to constrain all possible WIMP scenarios at once.
For this reason, it is of extreme importance that the LHC collaborations publish their results on simple, search-specific, models in such a way that they are recastable for any other model (as it is for cut-and-count analyses). In turn, theoreticians should keep working in close contact with experimentalists in order to maximise the utility of the simplified models tool-kit.
Finally, the use of (truncated) EFT should not be disregarded, since this is the most model independent approach and it is economical from the point of view of the reduced dimensionality of its parameter space.

\section*{Acknowledgements}

I thank Farinaldo Queiroz and Giorgio Arcadi for the invitation to write this contribution. I am grateful to Michael Duerr and Davide Racco for the many useful comments on this manuscript.

\bibliographystyle{JHEP}
\bibliography{lit}

%%%%%%%%%%%%%%%%%%%%%%
% End of vietnam.tex %
%%%%%%%%%%%%%%%%%%%%%%

\end{document}